\documentclass[a4paper, 11pt, twocolumn]{article}

\usepackage{inputenc}
\usepackage[T1]{fontenc}
\usepackage{amssymb}
\usepackage{hyperref}
\usepackage{apacite}
\usepackage{graphicx}
\usepackage{natbib} 
\usepackage{subfig}

\begin{document}

\title{Designing Touchscreen Menu Interfaces for In-Vehicle Infotainment Systems: the Effect of Depth and Breadth Trade-off and Task Types on  Visual-Manual Distraction}
\author{Nicolas Louveton$^1$ \and Rod McCall$^2$ \and Thomas Engel$^3$}
\date{
    $^1$Université de Poitiers, Université François-Rabelais de Tours, CNRS, Poitiers, France \\ \url{nicolas.louveton@univ-poitiers.fr}\\%
    $^2$Luxembourg Institute of Science and
Technology, Esch-sur-Alzette, Luxembourg \\\url{roderick.mccall@list.lu} \\
    $^2$University of Luxembourg, Luxembourg \\ \url{thomas.engel@uni.lu} \\
\bigskip
\today} 

\maketitle
     
\abstract{ \it
Multitasking with a touch screen user-interface while driving is known to impact negatively driving performance and safety. Literature shows that list scrolling interfaces generate more visual-manual distraction than structured menus and sequential navigation. Depth and breadth trade-offs for structured navigation have been studied. However, little is known on how secondary task characteristics interact with those trade-offs.
In this study, we make the hypothesis that both menu's depth and task complexity interact in generating visual-manual distraction.
Using a driving simulation setup, we collected telemetry and eye-tracking data to evaluate driving performance. Participants were multitasking with a mobile app, presenting a range of eight depth and breadth trade-offs under three types of secondary tasks, involving different cognitive operations (Systematic reading, Search for an item, Memorize items' state). 
The results confirm our hypothesis.
Systematic interaction with menu items generated a visual demand that increased with menu's depth, while visual demand reach an optimum  for Search and Memory tasks.
We discuss implications for design: In a multitasking context, display design effectiveness must be assessed while considering menu's layout but also cognitive processes involved. 
}

\section{Introduction}
\label{sec-intro}
In the last decade, in-car touchscreen displays have become prevalent to facilitate access to infotainment services, using either embedded systems or smartphone integration. It is well known that multitasking while driving decreases driving performance and safety~\citep{Alm1995, Young2007b, Treffner2004, Horberry2006a}; and this is  particularly true when interacting with touchscreen displays (see for example \citealt{kujala2013visual,kim2014evaluation}). For this reason, those in-car displays should be designed with safety constraints in mind \citep{green2008driver}.

Although, autonomous driving promises to relief driver's attention,  autonomy will likely be delivered gradually with a range of shared authority patterns \citep{J3216_202107}. Let alone, the fact that despite legislation, drivers frequently use touchscreen displays in the car \citep{Ahlstrom2020,makela2017naturalistic}. For those reasons, understanding how the design of in-car touch-screen displays impacts driving safety remains a key question.

\subsection{Navigating information and interruptibility}

Driving is fundamentally a visual-manual activity. For this reason, when the driver gets involved into a secondary task which requires visual and manual attentional resources (typically, texting on a mobile phone), driving safety decreases dramatically. Indeed, engaging in concurrent tasks which are plugged into the same attentional resources pool will generate interference and decrease overall performance \citep{wickens2008multiple}.

While it could be still valuable to access a range of infotainment features from the car (listening to radio or music, getting navigation or car related information, etc.), interactions must respect some constraints in order to guarantee a sufficient control of the car \citep{green2008driver}.  Those constraints imply the ability of the driver to switch relatively easily from the secondary task to the driving one by minimizing single glance fixation time; and favorite fast termination of the secondary task.

In other words, display design must support interruptibility \citep{Salvucci2009, salvucci2010reconstruction, burns2010importance} in the sense that the driver must not be locked-down into a given task. If it is not possible to achieve this, then this specific feature must be only available as the car is stopped. This idea is consistent with former results on higher performance in dual-tasking when the secondary task could be chunked in more manageable subtasks \citep{Janssen2012,LEE2019104}.

While specific tasks such as entering text with a visual keyboard are prohibited due to its high demand on visual-manual processes, it is still common to navigate through information using menus. Menus are displaying features or content as linear or grid lists. List items are navigated either through a screen scrolling, a paging or a hierarchy of categories.
Accelerating interactions with a menu is achieved by supporting the ability of the user to skip irrelevant content  \citep{hinckley2002quantitative}. This is possible when the user can predict where relevant items are in the menu {hick1952rate, hyman1953stimulus}). There are two ways of skipping through irrelevant information on a touchscreen display: kinetic scrolling and discrete navigation using hierarchies and pages.

\subsection{Discrete browsing offers a better interruptibility}

Due to smartphone and touchscreens popularity, kinetic scrolling became a mainstream way to browse through a list of items. Indeed, kinetic scrolling enable the user to increase the scrolling rate using accelerated fingers movement on the surface of the screen. This approach not only allows browsing a list at various speed but also requires a small precision in finger pointing as the whole screen is sensitive to directional fingers movements (see Fitts' law, \citealt{fitts1954information}). 

However, although this approach offers a familiar metaphor, transferring it to in-vehicle interfaces has been pointed out as a major source of distraction for drivers \citep{kujala2013visual}. Indeed, several studies showed an important adverse effect of using kinetic scrolling while driving \citep{kim2014evaluation, kujala2013visual, rydstrom2012comparison, lasch2012designing, kujala2013browsing}. While Fitt's law predicts that kinetic scrolling will be easier to operate, controlling an accelerated viewport on a continuous list do decrease multitasking performance. A possible explanation  \citep{brouwer2002perception, Bennett2013,louveton2016driving} is that accelerated movements are difficult to predict and control, increasing demand on visual-manual attention resources.

In contrast, literature showed that discrete browsing (parts of the items are presented on a screen which is moved thanks to a button or a gesture) generates less visual-manual distraction in driving context
\citep{lasch2012designing, kujala2013browsing}. While such a navigation pattern requires more actions from the user, it seems to relieve the driver from an accurate localization and control of the list widget.

Discrete information navigation is closely related to information hierarchies. Indeed, items are to be sorted some way over the different screens. The sorting criteria could be alphanumeric or semantic. Both way, the user is able to predict the location of the item he or she is seeking. When the number of item increases, items could be distributed either on fewer screens favoring breadth, or on more screens favoring depth (see also Figure \ref{fig:layouts}). This, in turn, rises the question of the best depth and breadth trade-off. Former studies showed that task completion time was following  a U-shape curve with a minimum favoring breadth (more items on a given screen) over depth \citep{miller1981depth, kiger1984depth, snowberry1983computer, landauer1985selection}. 

In the context of in-vehicle interfaces, Burnett et al. \citep{burnett2013menu} explored the impact of depth and breadth trade-off on driver's performance.
He used an occlusion protocol to measure visual attention and presented both structured (i.e., alphabetically ordered) and unstructured (i.e., random order) menus. They found that for structured menus' completion time and impact on visual attention was better for higher breadth (i.e., more items on a given screen); conversely, for unstructured menus they found a linear trend pointing to higher depth (i.e., more categories) to be better. These results are in line with those of \cite{cockburn2009predictive, landauer1985selection}.

\subsection{Present study}

Overall, it seems that in a driving context, touchscreen menus must be designed to allow discrete interactions, and favorite breadth over depth as they are likely to be structured in order to decrease their visual-manual distraction potential.

This literature review shows that kinetic scrolling is not appropriate for dual-task contexts such as driving. It also shows that a step-wise navigation process, distributing items over several screens, presents better properties to be use in a dual-task context. Finally, a studies point to a better performance of menu layouts favoring breadth over depth, particularly with predictable item locations. However, we found few studies investigating this trade-off in driving context, particularly involving eye-tracking measurements. Additionally, most of former literature considered list interfaces as support for a visual search task. However, the nature of the task that should be performed by the user might impact the optimal trade-off for a list interface.

In this driving simulator study, we will investigate the impact of  screen-wise navigation on driving safety, using eye-tracking measurements and testing a range of depth and breadth layouts. While performing a car following task, eight items were presented to participants under four different layouts (1 page x 8 items, 2x4, 4x2, and 8x1). Additionally, those four list layout were supporting three types of task: simple visual search, systematic interaction with each item, or systematic interaction with a memory component.

\section{Methodology}
\label{sec-meth}
\subsection{Participants}
\label{sec-2-1}

Twenty-eight participants took part in this study (20 males / 8 females). They were aged from 20 to 45 ($28.6 \pm 6.5$, m $\pm$ sd). We recruited participants who hold a driving licence for at least three years ($9.8 \pm 6.8$, m $\pm$ sd) and with a normal or corrected-to-normal vision. All participants completed and signed an informed consent at the beginning of the experiment.
   
\subsection{Experimental design}
\label{sec-2-2}

\begin{figure}[!h]
\begin{center}
  \includegraphics[width=1.0\columnwidth]{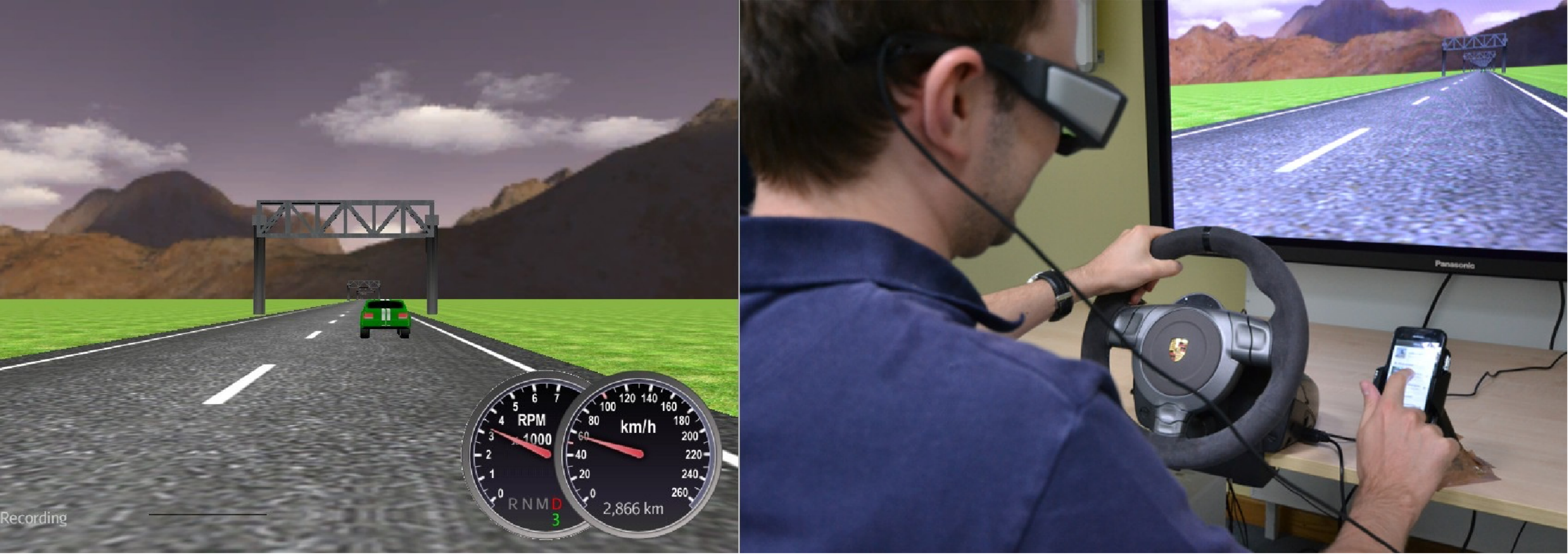}
  \caption{Experimental task (left) and set-up (right): In each of the three trials, participants were to follow the lead car while they interacted with a docked smartphone in the driving simulator.}~\label{fig:setup}
  \end{center}
\end{figure}

Participants were placed in a simulated driving environment (see Figure ~\ref{fig:setup}, left). In each trial, participants performed a car following task. They were to drive on a straight 5-km road and to follow a lead vehicle without overtaking it. They were driving on a two-lane seven-meter width road with grass on each side. Thirty-two bridges were positioned along the way to enhance speed perception. A trail vehicle was also following participants (constant speed, 50 km/h) and was visible in the rearview mirror. The speed of the lead vehicle changed 16 times during a trial, at regular intervals. The speed was selected randomly under a uniform distribution centred on $50$ km/h, $\pm$ 15 km/h. Thus, lead car velocity profile was randomized for each trial. Participants were told to follow the lead vehicle as close as possible without compromising their safety. 

While participants were performing the car following task, they interacted with a secondary task displayed on a smartphone, next to the steering wheel. Eight times in a trial, a new task was activated on the smartphone. Participants were alerted of the new task by a prior visual and auditory notification. The eight secondary task triggers were distributed equally on the road. To avoid the learning effect, a random shift of $\pm$ 100 meters was applied to all of them around their initial position.

\subsection{Secondary task}
\label{sec-2-3}

The secondary task was displayed through a smartphone application: A list of numbered items (eight in total) was presented on the left side of the screen, with a corresponding toggle button (On/Off state) on the right. For each item, a desired state is indicated with the item number (e.g., ``ITEM 1: ON'') and the task of the participant is to set the state of toggle buttons accordingly.

\begin{figure}[!h]
\begin{center}
  \includegraphics[scale=.4]{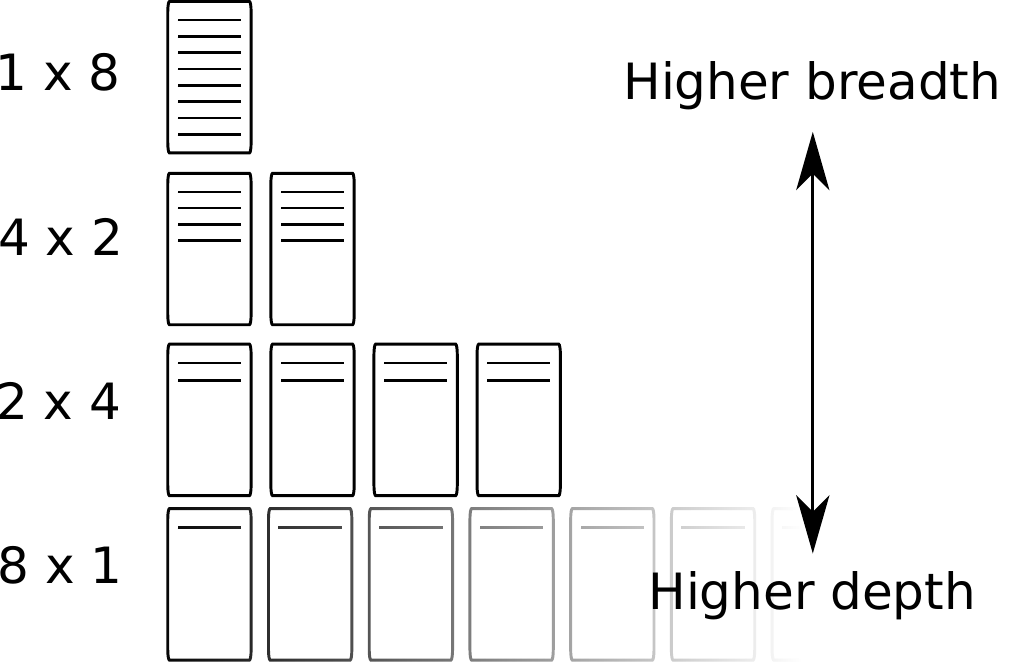}
  \caption{Schema describing the different layouts used in this experiment: For a constant menu length, higher breadth implies more items displayed on fewer screens, while higher depth implies fewer items displayed per screen but distributed across more screens. }~\label{fig:layouts}
  \end{center}
\end{figure}

Two within-subject experimental factors were manipulated. First, the layout of the interface with a range of depth and breadth trade-offs (see also Figure \ref{fig:layouts}): Participants had to interact with the eight items distributed evenly on one page (1 page x 8 items layout), two pages (2 x 4), four pages (4 x 2) and eight pages (8 x 1). A Next/Finish button at the bottom-right of the screen allowed participants to navigate through the menu list.

\begin{figure}[!h]
\begin{center}
  \includegraphics[scale=.2]{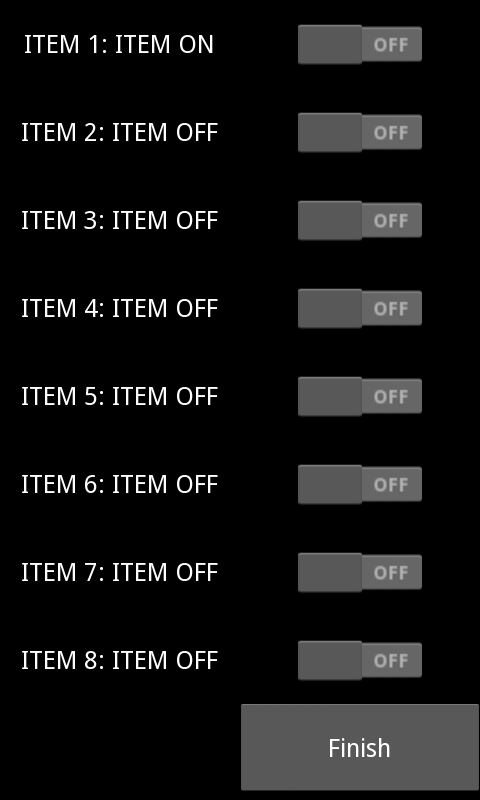}
  \includegraphics[scale=.2]{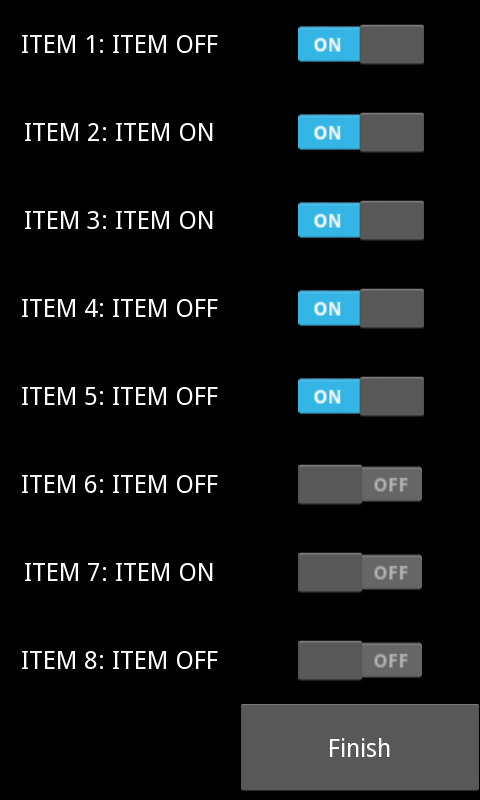}
  \includegraphics[scale=.2]{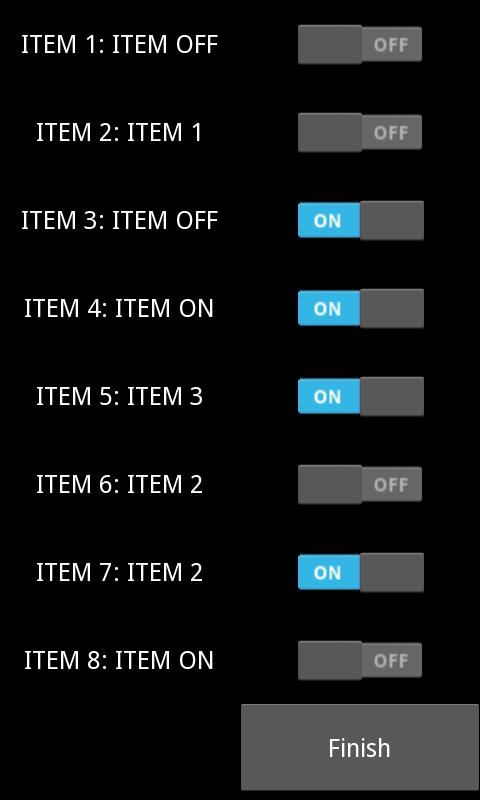}
  \caption{Representation of the three tasks used in this experiment under a one-page layout (see text for explanations).}~\label{fig:tasks}
  \end{center}
\end{figure}

We also manipulated the type of task (see Figure \ref{fig:tasks})) to be performed by participants. They had to perform three different tasks: \textbf{Search} for a specific item, \textbf{Systematic} reading of items and \textbf{Memory} consisting in memorizing items' state. In the Systematic task, participants had to read systematically each item's desired state and to set correctly the corresponding toggle button. In the Search task, participants had to visually search the item that should be tuned On and activate the corresponding button.  Finally, the Memorize task was similar to Systematic one except that some desired state were pointing to the state of a previous item (e.g., ``ITEM 5: ITEM2'', the state of item 5 must be equal to that of item 2). In total, four of those carry-over items were placed randomly in the list, referring to a randomly selected previous item.

\subsection{Apparatus}
\label{sec-2-4}

In this study, we used a low-cost driving simulator setup (See also \citealt{Jamson2010d}; for a discussion on low-cost simulator studies). Participants were placed in front of a Panasonic plasma screen of 165 cm diagonal, full HD resolution.  To control the simulated car, a Fanatec Porsche 911 GT2 steering wheel and pedal set were used. The simulator used an automatic transmission gear box. Additionally, to investigate eye fixation behavior, eye-tracking data were captured with a SMI Eye Tracking Glasses.

The 3D simulation engine used was OpenDS 2.5 \footnote{\url{http://www.opends.eu/home}} (\citealt{math2013opends} ; modified in order to obtain telemetry data in real time). The iGear platform \citep{avanesovtowards,louvetonaui13} was logging data from the simulator and the smartphone, and was in charge of triggering secondary task events.

The touchscreen device used in this study was a Samsung Galaxy S3 mini smartphone with four inches screen (HVGA $480\times800$ resolution). It runs Android 4.1 as operating system. The smartphone was placed on the right side of the steering wheel, with a standard dock (suction cup) for the smartphone (see Figure \ref{fig:setup}, right panel). The orientation of the smartphone's screen was adjustable, but not the position of the dock relative to the steering wheel.

\subsection{Procedure}
\label{sec-2-5}

Participants had first to fill out a profile questionnaire and an informed consent form. Next they were placed in the driving simulator setup with the possibility to adjust the seat and the pedals position.  Then, the test administrator explained the experimental task and gave participants the opportunity to practice each of the three secondary tasks. Finally, the eye-tracker was calibrated (three points calibration).

All participants completed a familiarization phase with the simulator. During this phase, they drove along a 1-km road without secondary task. Then they took part in the data collection phase consisting of three experimental trials: one for each type of secondary task (i.e., Search, Systematic, Memorize). Those three tasks were randomly assigned to  trials for each participant. After each trial, participants were asked to complete a full NASA-TLX questionnaire \citep{Hart2006}.

\subsection{Data analysis}
\label{sec-2-6}

Secondary task performance has been measured through completion time, success rate (i.e., number of correct items out of eight) and NASA-TLX scores. Concerning driving performance, we analyzed both longitudinal and lateral control. Longitudinal control was measured through average speed and CG Headway (CG standing for center of gravity). The CG Headway was calculated as the distance between the two cars' geometric center (also see SAE J2944 \citealt{j2944}). Lateral control was measured through the Standard Deviation of Lane Position (SDLP). It was calculated using the unbiased estimation of standard deviation applied to the lane position data (i.e., distance from the lane and car's geometric center). See also SDLP option A in \citep{j2944}.

Eye-tracking data were coded in a frame-by-frame fashion, using a region of interest for detecting whether the participant was looking at the device while driving. Saccades duration times were not included in the on-device glances classification. Gaze data was synchronized with the telemetry and application data using a signal sent by the iGear server when the participant is starting and finishing each trial.

Before calculating averages, we extracted from telemetry and eye-tracking data the slices of time corresponding to when a secondary task was in execution. Parametric statistical tests (generalized linear regression model) were used whenever the validity conditions were met. Otherwise, we used non-parametric tests. Post-hoc tests were performed using pair-wise two-sample tests with a Bonferronni correction. Data analysis was carried out with Pandas Python library~\citep{pandas} and R~\citep{R}.

\section{Results}
\label{sec-resu}
\subsection{Secondary task performance}

Overall, success rate was very high across the different conditions, with a mean scores ranging from 7.2 to 8 (i.e., maximum eight correct items out of eight). This shows that, despite different difficulty levels (see below, NASA-TLX results and completion time), participants managed to keep a high performance level.

\subsubsection{NASA TLX}

\begin{figure*}[!ht]
  \subfloat[Memory \label{subfig:tlx-memory}]{%
    \includegraphics[scale=.35]{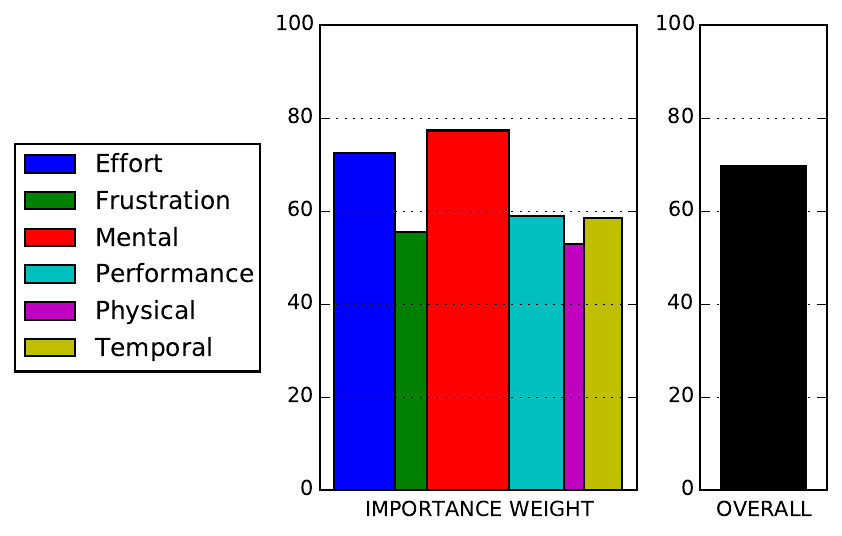}
  }
\hfill
\subfloat[Search \label{subfig:tlx-search}]{%
    \includegraphics[scale=.35]{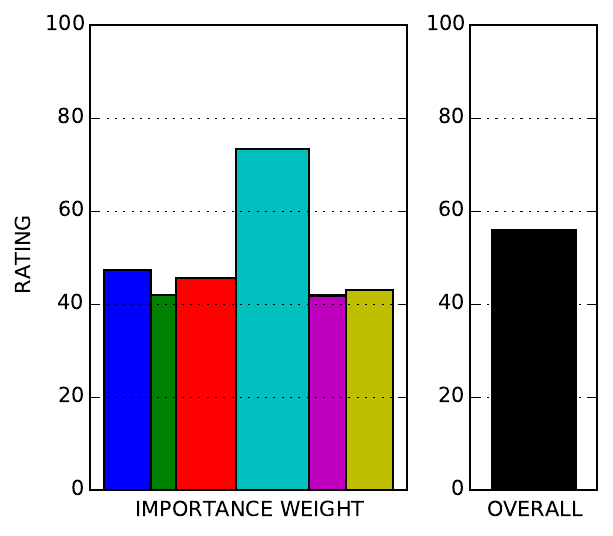}
  }
\hfill
  \subfloat[Systematic \label{subfig:tlx-systematic}]{%
    \includegraphics[scale=.35]{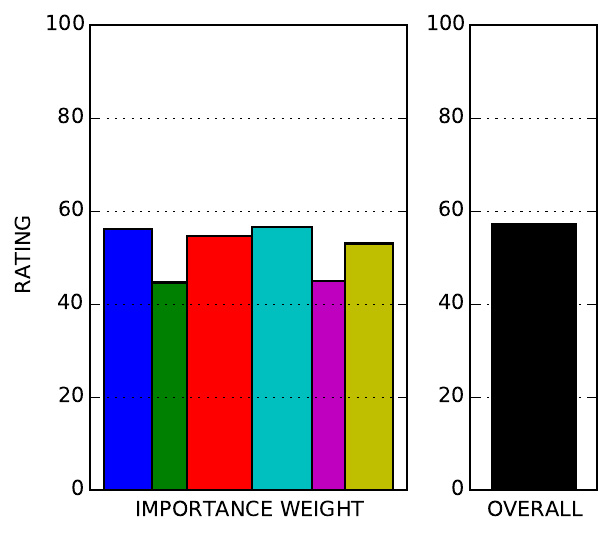}
  }

  \caption{NASA-TLX rankings for the three tasks: The global workload estimation was higher for the Memory condition, while the two other tasks were much closer to each other. However, the weights of the \textit{Effort} and \textit{Mental} scales were stronger in Systematic than in Search, which in turns presented a higher weight for the \textit{Performance} compoenent. }
  \label{fig:nasa-tlx}
\end{figure*}

Results showed (see Figure~\ref{fig:nasa-tlx}) that global workload score was the highest for Memory task (69.8, sd = 9.8) followed by the Systematic (57.1, sd = 11.7) and Search (55.8, sd = 13.5). A Friedman test revealed significant differences ($\chi^2(2)=20.8, p <.001$). Post-hoc comparisons (pair-wise Wilcoxon Sign test with Bonferroni correction) showed that Memory task differed significantly from both Search and Systematic ones ($ps < .001$). Finally, no differences were found between Search and Systematic.

\subsubsection{Completion Time}

\begin{figure}
\begin{center}
  \includegraphics[width=1.0\columnwidth]{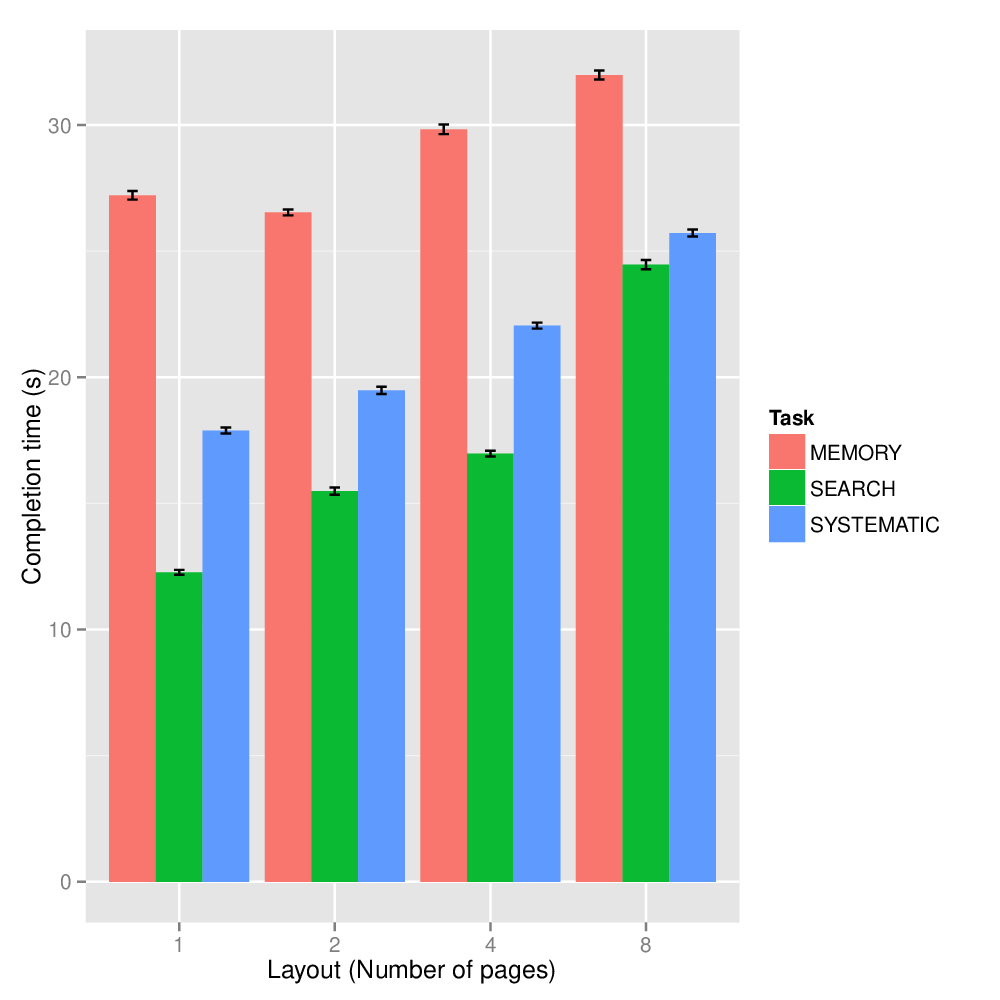}
  \caption{Completion time increases gradually with the number of pages to navigate through. In the Memory condition, this pattern is less clear and participants took a much longer time to complete the task than in other conditions.}
  ~\label{fig:res-ct}
  \end{center}
\end{figure}

A two-way repeated measures ANOVA showed an effect of Layout ($F(3,238)=99.2, p < .001$), Task ($F(2,238)=36.5, p < .001$) and of the interaction of both factors ($F(6,238)=5.5, p < .001$). Completion time (see also Figure \ref{fig:res-ct}) was the lowest for Search task (17.1, sd = 7.9), followed by Systematic (21.1, sd = 6.7) and Memory ones (28.6, sd = 7.3). Also, completion time is increasing gradually with the number of pages: the lowest was observed with the one-page layout condition (19.1, sd = 9), followed by the two-page (20.5, sd = 8), four-page (22.3, sd = 8.2) and finally eight-page ones (27.1, sd = 7.5).

A post-hoc analysis (pair-wise t-test with Bonferroni correction) showed Memory task to be significantly different from Search and Systematic ($ps < .01$), while Systematic and Search were never found significantly different. When looking at the interaction between the two factors, we found that in the Memory condition, only the two- and eight-page layouts were found different ($p < .05$). In the Systematic condition, only the comparison between the one- and eight-page layouts was found significant ($p < .001$). Finally, in the Search condition, the one-page layout was found different from each of the two-, four- and eight-page layouts ($ps < .05$).

\subsection{Telemetry}

\subsubsection{Speed}

\begin{figure*}
  \includegraphics[scale=.3]{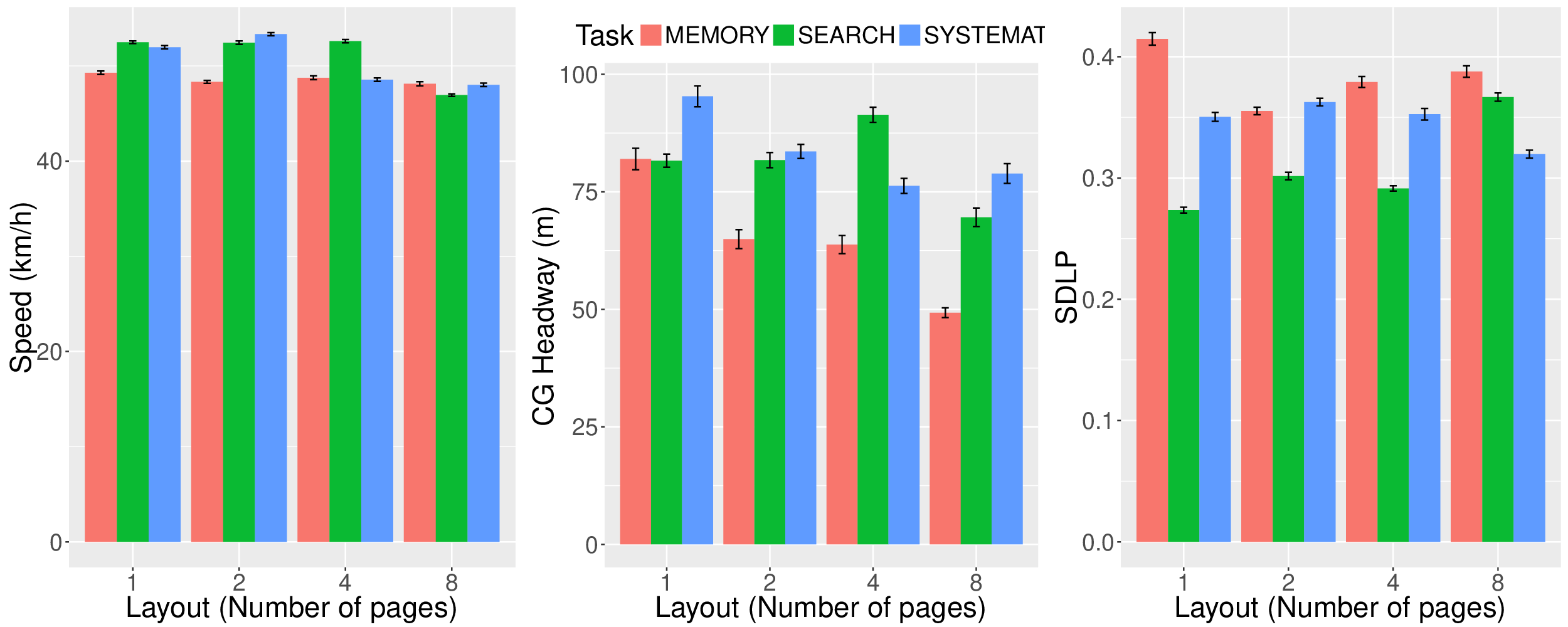}
  \caption{Driving behavior measured in the different experimental conditions}
  \label{fig:telemetry}
\end{figure*}

A two-way repeated measures ANOVA showed an effect of Layout ($F(3,238)=5.4, p < .01$), Task ($F(2,238)=3.4, p < .05$) and of the interaction of both factors ($F(6,238)=2.4, p < .05$). Mean speed (see also Figure \ref{fig:telemetry}, left) was the lowest for Memory task (48.7, sd = 8) followed by Systematic (50.5, sd = 8.4) and Search (51.3, sd = 8.2).  Mean speed was also the lowest for the eight-page layout (47.6, sd = 6.9), followed by the four-page (50.1, sd = 8.8), two-page (51.4, 8.4) and one-page ones (51.2, sd = 8.2). A post-hoc analysis  showed a significant difference only between the two-page layout in the Systematic condition and the eight-page layout in the Search one.

\subsubsection{CG Headway}

A two-way repeated measures ANOVA revealed only an effect of the Layout factor ($F(3,238)=2.8, p < .05$). Neither the Task factor was not found significant ($p = .4$) nor was the $Layout \times Task$ interaction ($p = .2$). When comparing CG Headway (Figure \ref{fig:telemetry}, middle) in different task conditions, we observed that the average distance was the smallest for Memory (66.4, sd = 89.7), followed by Search (81.6, sd = 79.9) and Systematic (83.5, sd = 86.2). When comparing mean CG Headway across page layouts, we observed that the smallest distance was found in the eight-page layout (66.8, sd = 75.6), followed by the two-page (76.8, sd = 84.3), four-page (78.5, sd = 80.8) and one-page ones (86.1, 97.1). A post-hoc analysis on the Layout factor did not evidence any significant differences.

\subsubsection{SDLP}

A two-way repeated measures ANOVA showed an effect of Layout ($F(3,238)=3.2, p < .05$) and of the  $Layout \times Task$ interaction ($F(6,238)=2.5, p < .05$). However, we did not find an effect of the Task factor ($p = .07$). Values of SDLP (Figure \ref{fig:telemetry}, right) were very close to each other in the different task conditions: Memory (.38, sd = .20), Search (.31, sd = .14) and Systematic (.35, sd = .18). Across the different layouts, SDLP measurements were very close to each others (ranging from .34 to .36). None of the post-hoc analyses revealed significant differences.

\subsection{Eye fixations}

\subsubsection{Number of fixations (NF)}

A two-way repeated measures ANOVA showed an effect of Layout ($F(3,174)=30.6, p < .001$), Task ($F(2,174)=5.5, p < .01$) and of the interaction of both factors ($F(6,174)=2.7, p < .05$).
Number of fixations (see also Figure \ref{fig:fixations}, left) was the highest for Memory task (11.5, sd = 15.6), followed by Systematic (5.3, sd = 5) and Search (4.9, sd = 4). Also, NF was the highest for the eight-page layout (10, sd = 10.4) followed by the four-page (6.9, sd = 7), two-page (6.3, sd = 12.7) and one-page (6.2, sd = 9.2) layouts. The post-hoc analysis did not evidence any significant differences.

\subsubsection{Mean single fixation duration (SFD)}

A two-way repeated measures ANOVA showed an effect of Task ($F(2,174)=6.1, p < .01$) and of the interaction of both factors ($F(6,174)=2.8, p < .05$). The effect of Layout factor was not found to be significant ($p = .3$). The mean single fixation duration (Figure \ref{fig:fixations}, middle) was the highest for Memory task (1.7, sd = 1.7), followed by Systematic (1.5, sd = 1.7) and Search (1.1, sd = 1.1). Also, SFD was the highest for the eight-page layout (1.7, sd = 1.7) followed by the two-page (1.4, sd = 1.7), four-page (1.3, sd = 1.6) and one-page (1.3, sd = 1.2) layouts. The post-hoc analysis did not evidence any significant differences.

\subsubsection{Cumulative fixation duration (CFD)}

\begin{figure*}
  \includegraphics[scale=.3]{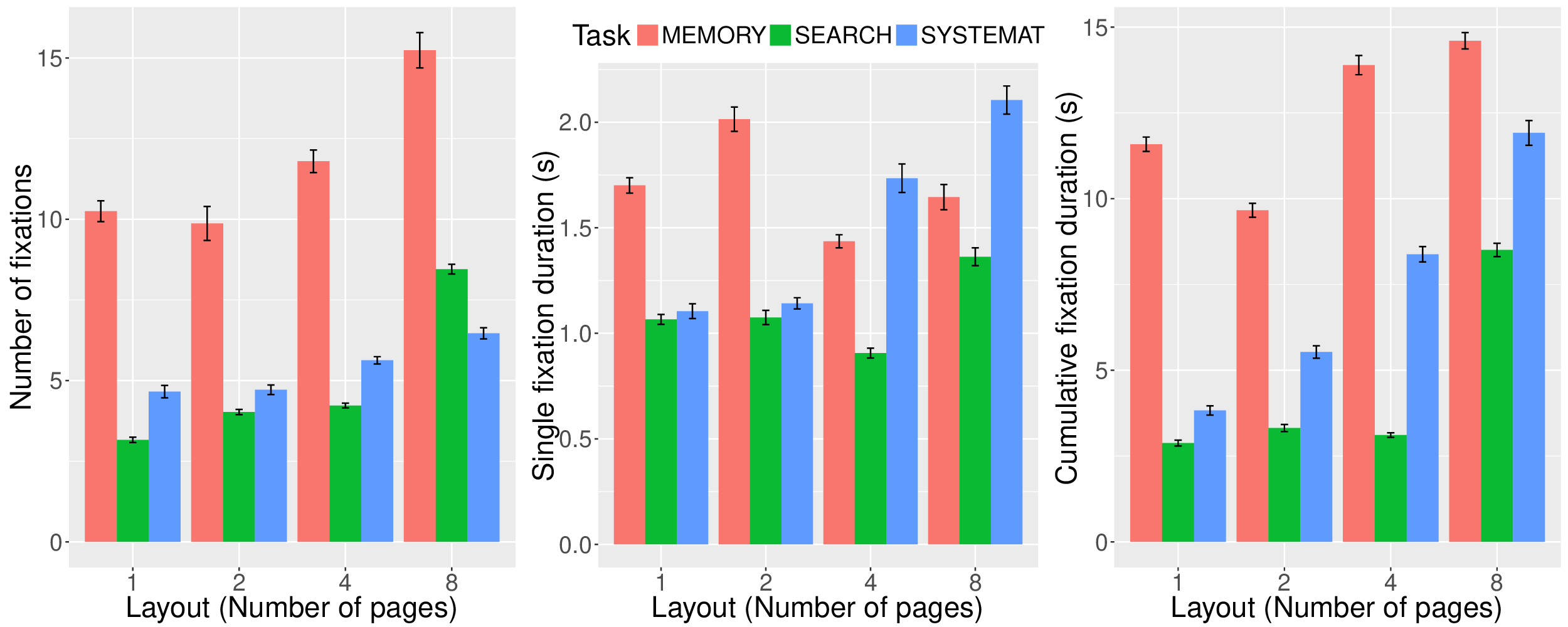}
  \caption{Fixation behavior under different conditions of Task and Layout: Number of fixations (NF), Single fixation duration (SFD)
and Cumulative fixation duration (CFD).}
  \label{fig:fixations}
\end{figure*}

A two-way repeated measures ANOVA showed an effect of Layout ($F(3,174)=23.2, p < .001$), Task ($F(2,174)=13.2, p < .001$) and of the interaction of both factors ($F(6,174)=3.3, p < .01$).
The cumulative fixation duration (Figure \ref{fig:fixations}, right) was the highest for Memory task (12.2, sd = 8.1), followed by Systematic (7.3, sd = 7.8) and Search (4.3, sd = 4.7). Also, CFD was the highest for the eight-page layout (11.5, sd = 8.2) followed by the four-page (8.0, sd = 7.9), one-page (6.3, sd = 7.1) and two-page (6.2, sd = 6.8) layouts.
When comparing layouts, post-hoc analysis revealed that for the Search task, the four-page layout was significantly different from the eight-page one ($p < .05$). The same analysis did not reveal significant differences for Memory and Systematic. When comparing tasks across each layout level, post-hoc analysis revealed that for the one-, two- and four-page layouts, Memory and Search tasks were significantly different ($p < .01$). 

\subsubsection{Visual demand}

\begin{figure}[!h]
\begin{center}
  \includegraphics[width=1.0\columnwidth]{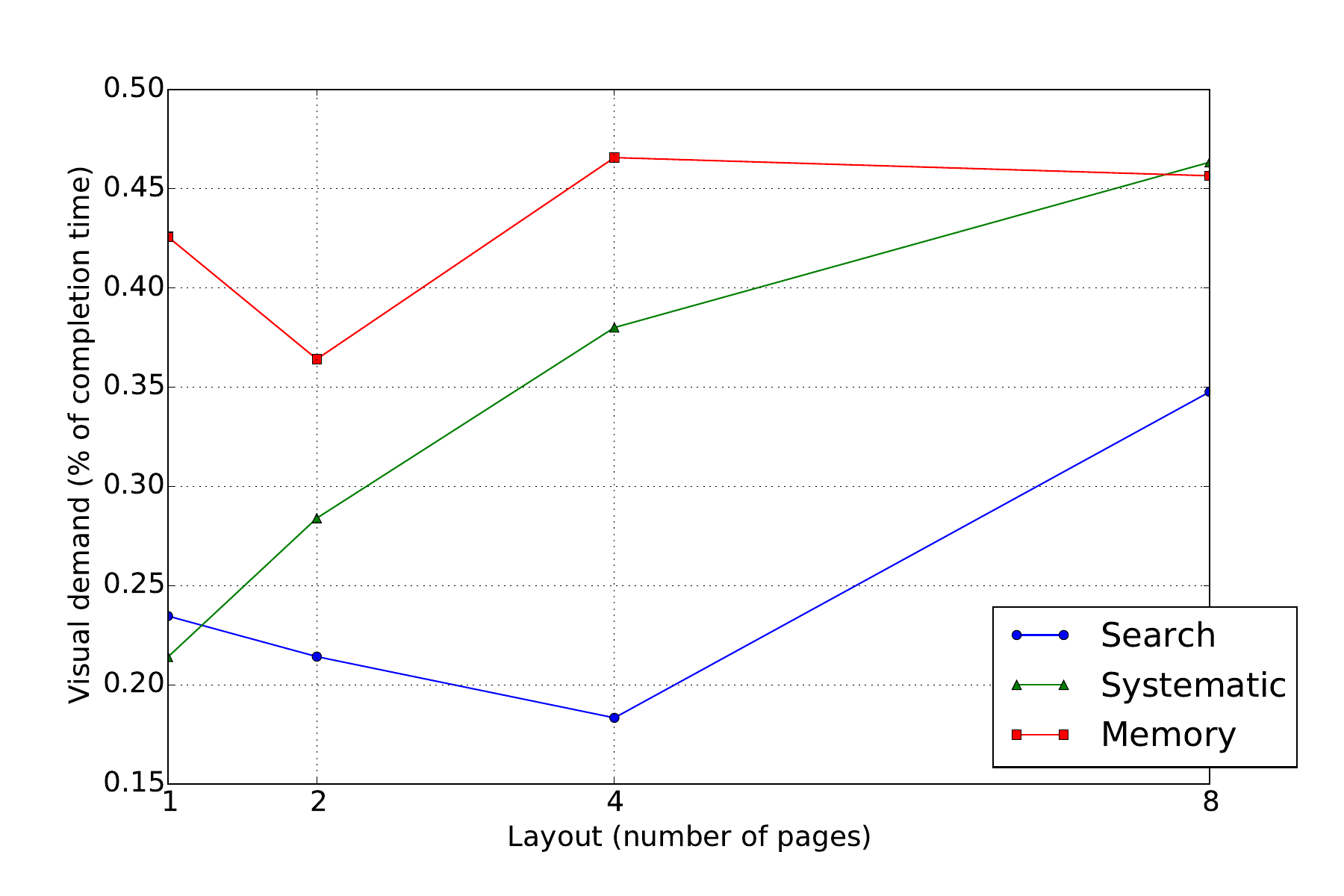}
  \caption{Time-sharing when multitasking: the highest the percentage, the lowest the visual interruptibility is. }~\label{fig:dis-tradeoffs}
  \end{center}
\end{figure}

In order to better understand how visual attention is distributed across the primary and secondary tasks, we calculated the percentage of time dwelling on the application against the time to complete the task (See Figure \ref{fig:dis-tradeoffs}). This metric is assumed to reflect the flexibility with which the user could interrupt the secondary task in order to focus on driving: higher percentages refer to a high visual demand with a low capacity to switch back to the driving task, while lower percentages refer to a low visual demand and a higher flexibility.

This figure shows that for the Systematic task there was no optimum, instead the visual demand increased continuously with the number of pages. Concerning Memory task, we found an optimum value on the two-page layout, while for Search we found an optimum on the four-page layout. Also, we found that for Memory task visual demand was always higher, although there were no additional visual operations except probably in visual working memory.

\section{Discussion}
\label{sec-disc}
We showed in a state of the art that menu navigation was the most appropriate way to display list information in an in-vehicle interface. We also showed that most of the studies pointed to breadth being preferred over depth in page-based navigation design \citep{landauer1985selection, cockburn2009predictive,burnett2013menu}. However, few studies investigated this trade-off in driving context (but see \citealt{burnett2013menu}) and involving different types of task. We reported results from a driving simulator and eye-tracking study which investigated how the depth and breadth trade-off varied with the type of task and how it impacted dual-task performance.

\subsection{Dual-task performance}

While success rate was high for each task and layout condition, Memory seemed to be the most demanding task when looking at completion times and NASA-TLX scores. This is not too surprising given that the Memory task was including a ``cognitive'' aspect in addition to visual-manual distraction also present in other tasks. 

Completion time also increased with the number of pages: This tendency was particularly visible for Systematic and Search tasks, while it did not appear as clearly for Memory task. This linear trend reproduces earlier results on structured lists \citep{landauer1985selection, cockburn2009predictive,burnett2013menu}. Because it involved memorizing randomly selected former items, the Memory task presented results that might be closer to those obtained for unstructured lists (non-predictable item location).

Concerning driving performance, when participants were engaged in a Memory task, we observed that they drove at a slower speed, the lateral control of their vehicle was more variable and the distance to the lead vehicle was smaller. Those findings seem to indicate a higher difficulty to control the vehicle in this situation. Other analyses did not help to distinguish between experimental conditions. No clear trade-off as been identified on the basis of telemetry results.

\subsection{Impact on attention}

Results showed that concerning Memory task, the number of fixations is always higher than in other conditions and so it is for the single-fixation duration and the cumulative fixations' duration (although we did not always observe significant differences). 

Considering Systematic task, we observed that the three measurements are increasing with the number of pages to navigate through, while this pattern was less clear for Memory and Search.

Finally, concerning the Search task, time spent on the application is relatively stable across the one-, two- or four-page layouts while it increases abruptly for the eight-page layout (i.e., one item per page). This seems to indicate that in a visual search task, parallel processing could occur no matter the size of the information chunk, although forcing serial search decreases clearly the performance.

We computed a visual demand index based on eye-tracking and usage data (See Fig. \ref{figures/tradeoff}). Our finding shows that there were no optima for the Systematic reading task. In contrast, the higher the number of items, the higher the visual demand. This result suggest that systematic interaction tasks require distributing the menu content across more separate screens. On the other side, the Search task seemed to display an optimal depth and breadth hierarchy in terms of visual demand. This suggest that human are able to skip unnecessary information easily. For this reason, too deep or too large menus might impart negatively this ability. Finally, we also observe an optimal depth and breadth tradeoff with the Memory task, despite it concerns each item on the list, suggesting that memorization tasks might behave similarly to search task in visuospatial memory. 

Systematic task there was no optimum, instead the visual demand increased continuously with the number of pages. Concerning Memory task, we found an optimum value on the two-page layout, while for Search we found an optimum on the four-page layout. Also, we found that for Memory task visual demand was always higher, although there were no additional visual operations except probably in visual working memory.

While all combinations were tested, this study did not include numerous layouts, making it difficult to draw a definitive conclusion on a specific  trade-off value. However, we could conclude that for a systematic interaction task, visual demand increases with the number of pages while optimum could be found for Memory and Search tasks. Consistently with literature, this optimum is in favor of breadth over depth.

\section{Conclusion}
\label{sec-conc}
In safety-critical contexts such as driving, multitasking constraints should be carefully analyzed in order to design interfaces with appropriate trade-offs in terms of visual demand. Indeed, it is necessary to organize the information space in a way that reduces visual demand and increases the driver's ability to keep his/her attention on the road. In this study, we assessed the impact of four list interface layouts and three different types of task on vehicle control, secondary task performance and eyes fixations. Our results were consistent with former literature. We conclude that when a user should interact with all items of a list, the visual demand increases continuously with the depth of the menu, while for visual search and memorization tasks an optimum could be found with a preference for breadth over depth.

\section{Acknowledgements}

The iGear project was funded by the National Research Fund, Luxembourg
(Project code: 11/IS/1204159). We also thank other colleagues and
students from the iGear team.

\bibliographystyle{apacite} 
 \bibliography{refs}

\end{document}